\documentclass{aa}
\usepackage{epsfig}
\newcommand{\logra}{$\log \rm L_{\rm X}/\rm L_{\rm bol}$}
\newcommand{\vmi}{V$-$I}
\newcommand{\bmv}{B$-$V}
\newcommand{\bmo}{B$-$V$_0$}

\newcommand{\teff}{T$_{\rm eff}$}

\newcommand{\vmo}{V$-\rm I_0$}

\newcommand{\ewk}{K~{\sc i}~EWs}
\newcommand{\ewl}{Li~{\sc i}~EWs}

\begin{document}

\title{On the dispersion in lithium and potassium\\ among late-type stars 
in young clusters:\\
\object{IC~2602}
\thanks{Based on observations
carried out at the European Southern Observatory, La Silla, Chile}}

\author{Sofia Randich \inst{1}}

\offprints{S. Randich, \email{randich@arcetri.astro.it}}

\institute
{Osservatorio Astrofisico di Arcetri, Largo Fermi 5, I-50125
Firenze, Italy}

\titlerunning{On the dispersion in lithium}

\date{Received / Accepted} 

\abstract{
We have measured
the equivalent width (EW) of the K~{\sc i}~7699~\AA~line
in a sample of G and K--type
members of the $\sim 35$~Myr old cluster
IC~2602 for which a dispersion in Li EWs had been reported by previous
studies. Active cluster stars with
$0.75 \la$~\bmo~$\la 1$ are characterized by a dispersion in the EW of the
K~{\sc i}~7699~\AA,
while earlier and later--type stars do not show any significant
scatter. Cluster stars at all colors show potassium EW excesses
with respect to field inactive stars; furthermore, a statistically significant 
relationship is found between
differential potassium EWs and \logra~ratios, indicating that the EWs of
the potassium feature are altered by activity. 
Our results suggest that the dispersion in Li EWs observed
among cluster stars later than \bmo~$\sim 1$ cannot be fully explained
by the effects of activity. No final conclusion can instead be drawn for
earlier--type stars.
}
\maketitle
\section{Introduction}
The existence of
a star-to-star scatter in Li abundances among otherwise similar
late--type members
of the 120~Myr old \object{Pleiades} cluster was first reported by
Duncan \& Jones (\cite{dj83}) and Butler et al. (\cite{but87});
the dispersion 
was subsequently confirmed by several studies
and additional observational constraints were put on it
based on larger samples (Soderblom et al. \cite{sod93a}, Garc\'\i a
L\'opez et al. \cite{gar94}, Jones et al. \cite{jon96}).
A star-to-star
scatter in Li has also been reported for other clusters both younger and older
than the Pleiades, such as \object{Alpha Per} 
(Balachandran et al. \cite{bala96},
Randich et al. \cite {R98}),
IC~2602 (Randich et al. \cite{R97} --hereafter R97, Randich et al. \cite
{R01} --hereafter R01),
\object{IC~4665} (Mart\'\i n \& Montes \cite{mart97}), \object{M~34}
(Jones et al. \cite{jon97}),
\object{NGC~6475} (James \& Jeffries \cite{jj97}, James et al. \cite{jam00},
Randich et al. \cite{R00}). The scatter has disappeared by the Hyades
age (600~Myr). 

The detection of a dispersion is one of
the most puzzling results within the context of the so--called
Li problem (e.g. Jeffries
\cite{jeff00} and references therein), being in strong contradiction
with the predictions of ``standard models" of stellar evolution; 
standard models
incorporate convection only as a mixing mechanism and predict that
the amount of Li depletion should depend on mass, age,
and metallicity (or
chemical composition) only; no differences in Li abundances are indeed
expected for co-eval, otherwise similar stars in clusters.
We mention in passing that the spread
in Li in the Pleiades was initially ascribed to a large spread in age;
Soderblom et al. (\cite{sod93a},\cite{sod93b}), however, convincingly
showed that this is unlikely the case. 
Also note that the hypothesis that errors in stellar parameters, in particular
effective temperature, might be a source
of dispersion is rather unlikely, since a large scatter is present in
the EW vs. color diagrams, i.e., stars with the same color (and presumably
mass) in the same cluster have different Li EWs.
\begin{figure*}
\psfig{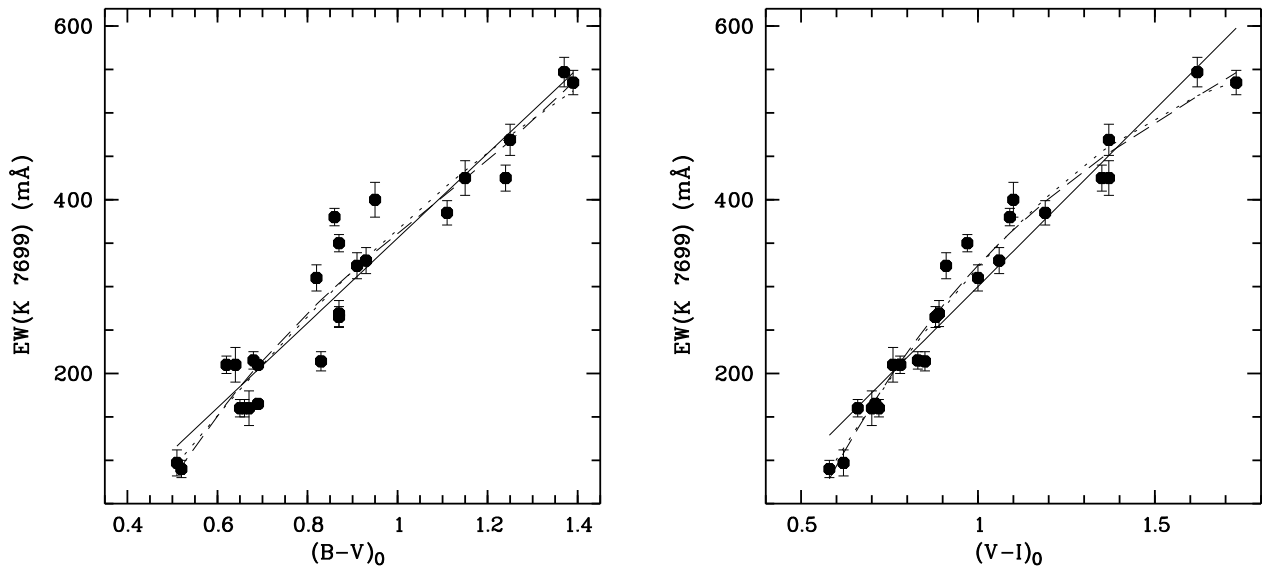}
\vspace{-2cm}
\caption{K~{\sc i}~7699~\AA~equivalent width vs. dereddened
stellar color
(\bmo~and \vmo~ in the left- and right-hand panels, respectively) for
our sample stars. The three curves indicate polynomial regressions
of grade 1 (solid), 2 (dotted), 3 (dashed).}
\end{figure*}
\begin{figure*}
\psfig{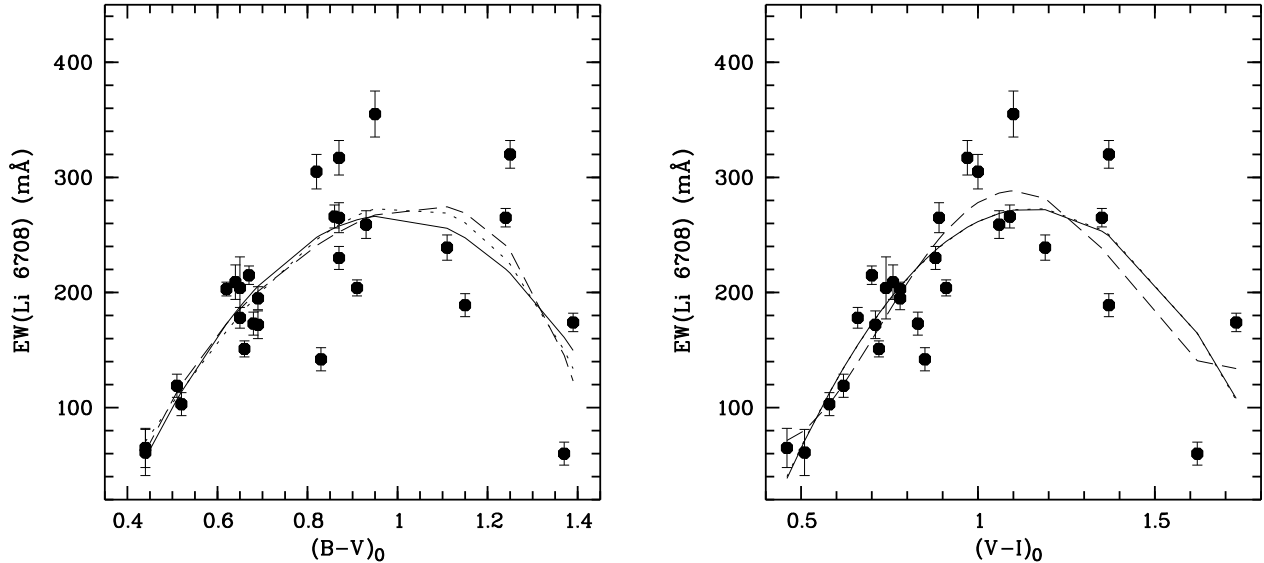}
\vspace{-2cm}
\caption{Same as Fig.~1, but the equivalent width of the Li~{\sc i}~6708~\AA
~line is plotted as a function of the two stellar colors. Grade 2 (solid),
3 (dotted), and 4 (dashed) polynomial regressions were carried out for this
line.}
\end{figure*}
\begin{figure*}
\psfig{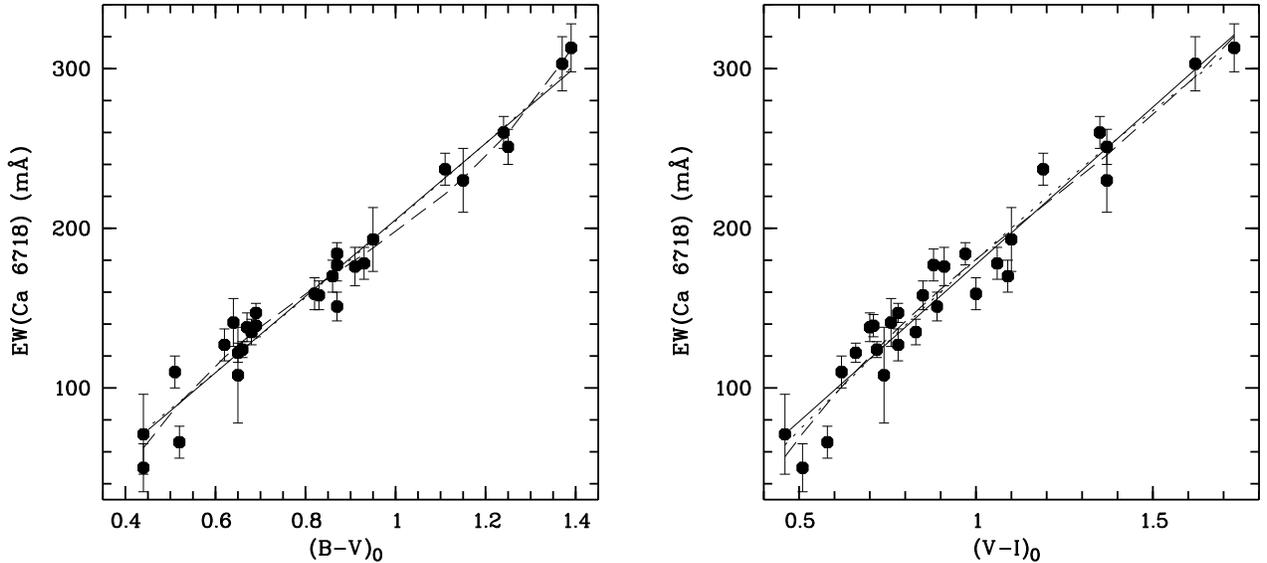}
\vspace{-2cm}
\caption{Same as Fig.~1, 
but the EWs of the Ca~{\sc i}~6718~\AA~line are shown.}
\end{figure*}
Under the assumption that the observed scatter in Li reflects
a {\it real} scatter in abundances, much work has been done on theoretical
grounds in order to explain it; extra-mixing processes
and/or mechanisms able to inhibit Li destruction were introduced in
the models. Driven by the observed Li--rotation relationship,
most of the models included rotation
and/or angular momentum loss
as an additional crucial parameter determining the amount of Li depletion
(e.g., Mart\'\i n \& Claret \cite{mart96},
Pinsonneault \cite{pins97} and references therein): according to these
models, different rotation rates or rotational histories would lead to
different Li abundances. None of the models so far elaborated
is able to quantitatively explain the observed features.

In the last few years there has been a growing interest in investigating 
whether and to which extent
the observed scatter in the EWs of
the Li~{\sc i}~6708~\AA~resonance doublet among young cluster stars
truly reflects a dispersion
in abundances or is rather due to other effects which
could affect the formation of the Li~line. More specifically,
the suggestion has been made that chromospheric
activity (including the presence of chromospheres as well as
surface inhomogeneities such as
spots and plages) may affect the formation of the Li line
and be the reason for the observed
spread in EWs, since, for a given abundance and temperature, more active
stars would have larger EWs. In this case, the dispersion would not
witness a dispersion in abundances and
the high-Li -- high-rotation relationship would not be a direct
relationship, but it would be the consequence of the
fact that high rotators are characterized by high
activity levels and thus their Li EWs are more affected by activity.

Various ways of addressing this issue exist:
we focus here on the simultaneous measurement of the 
Li~{\sc i}~6708~\AA~and the K~{\sc i}~7699~\AA~resonance features. 
The excitation potential of the potassium line and, more in general,
its formation conditions, are very
similar to those of the Li doublet; any line formation effect that
alters the Li line should also affect the K~{\sc i}~line and viceversa.
Potassium is not destroyed in stars and
star-to-star differences in its abundance among 
members of the same cluster are not expected: therefore, 
the detection of a spread in the EW of the K~{\sc i} line may provide
an indication that the scatter in Li EWs does not
necessarily imply a scatter in abundances. 

The effect of activity on the Li~{\sc i}~ and K~{\sc i}~lines
has been quantitatively studied
by Stuik et al. (\cite{stu97}), who modeled the atmospheric stratifications of
various combinations of spots and plages and investigated
how those stratifications
affect lithium and potassium line formation. They showed that the two
alkali lines are not sensitive to chromospheres, but they are
affected by the presence of plages and spots, which
also significantly
alter broad-band stellar colors and thus the observed alkali EW vs. color
diagrams. Stuik et al. were not able
to reproduce the Pleiades observational pattern
and concluded that, whereas it is not
easy to demonstrate that the dispersion in \ewk~(and \ewl)
is really and completely due to stellar activity, the presence of a
scatter in potassium EWs constitutes
a warning against interpreting the spread
in Li as due to a real spread in abundances.

A few additional studies on this topic were recently carried out:
Jeffries (\cite{jeff99})
using new data and old data from the literature, simultaneously monitored
the strengths of the Li~{\sc i} and K~{\sc i} lines and 
H$\alpha$ in a sample of Pleiades
K-type stars to search for variability of the line
strengths which would witness the presence of large-scale atmospheric
inhomogeneities. He detected no variability of \ewl~
on one year timescales, possible variability on 10 years timescales, and
only 20--30 \% variability in chromospheric activity; he confirmed the
presence of a dispersion in potassium and a correlation between Li and K line
strengths and rotation and activity. Similarly to Stuik et al. (\cite{stu97}),
he concluded
that the dispersion in \ewk~must be explained before definitively
accepting that the dispersion in Li is due to a genuine dispersion
in abundances.
King et al. (\cite{king00}) instead,
based on a new analysis of Pleiades data 
from the literature, found that an excess in Li abundance
correlates with an excess in the potassium EW
and activity and concluded, more firmly than the other studies, that activity
is, at least in part, the reason for the dispersion in Li EWs.
Finally, Barrado y Navascu\'es et al. (\cite{bar01}) presented an analytical
model to investigate the effect of stellar surface inhomogeneities
on the Li~{\sc i} and K~{\sc i} features (plus the Na~{\sc i}~5896~\AA~feature)
and compared their predicted EWs with the observed EWs in the Pleiades.
They concluded, that activity can explain part of the dispersion, but
it cannot fully account for it. The issue therefore is far from being settled.

So far, no studies of the potassium feature among late--type stars
in other young clusters have been carried out. In this paper we present
new potassium data for the 35~Myr old IC~2602; as mentioned above, a dispersion
in Li was detected among its late--type members by R97 and R01, although
the scatter seems narrower than in the Pleiades. With the present study
we wish to address the obvious questions whether IC~2602 stars are also 
characterized by a dispersion in \ewk~and whether there is a correlation
between potassium EWs and activity. A positive answer to these questions
would provide an additional hint that activity do affect the formation
of the alkali lines.
\begin{table*}
\centering
\caption{Sample stars and measured equivalent widths.}
\begin{tabular}{lccccc}
 & &  & & &\\ \hline\hline
 & &  & & &\\
star & \bmo & \vmo & EW~(Li~{\sc i}~$\lambda$~6708~\AA) & EW~(K~{\sc i}~$\lambda$~7699~\AA) &EW~(Ca~{\sc i}~$\lambda$~6718~\AA) \\
 & & & (m\AA) & (m\AA) & (m\AA) \\
 & &  & & &\\
 W~79 & 0.79 & 0.81 & 142$\pm$  5 & 214 $\pm$ 11 & 158 $\pm 9$\\
 R~1  & 0.87 & 0.87 & 204$\pm$  7 & 324 $\pm$ 15 & 176 $\pm 12$ \\
 R~3  & 0.83 & 0.85 & 265$\pm$ 13 & 269 $\pm$ 15 & 151 $\pm  9$\\
 R~8  & 0.61 & 0.62 & 178$\pm$  9 & 160 $\pm$ 10 & 122 $\pm  6$\\
 R~14 & 0.83 & 0.84 & 230$\pm$ 10 & 265 $\pm$ 12 & 177 $\pm 10$\\
 R~15 & 0.89 & 1.02 & 255$\pm$ 15 & 325 $\pm$ 15 & 180 $\pm 12$\\
 R~21 & 0.47 & 0.58 & 119$\pm$ 10 &  97 $\pm$ 15 & 110 $\pm 10$ \\
 R~29 & 1.07 & 1.15 & 239$\pm$ 11 & 385 $\pm$ 14 & 237 $\pm 10$ \\
 R~35 & 0.63 & 0.66 & 215$\pm$  8 & 160 $\pm$ 20 & 138 $\pm  9$ \\
 R~43 & 0.95 & 1.10 & 355$\pm$ 20 & 400 $\pm$ 20 & 193 $\pm 20$\\
 R~45 & 0.62 & 0.68 & 151$\pm$  7 & 160 $\pm$ 10 & 124 $\pm  5$\\
 R~54 & 1.11 & 1.33 & 189$\pm$ 10 & 425 $\pm$ 20 & 230 $\pm 20$ \\
 R~59 & 0.78 & 0.96 & 305$\pm$ 15 & 310 $\pm$ 15 & 159 $\pm 10$\\
 R~66 & 0.64 & 0.79 & 173$\pm$ 10 & 215 $\pm$ 10 & 135 $\pm  8$ \\
 R~68 & 0.82 & 1.05 & 266$\pm$ 10 & 380 $\pm$ 10 & 170 $\pm 10$ \\
 R~70 & 0.64 & 0.67 & 172$\pm$ 12 & 165 $\pm$  5 & 139 $\pm  7$ \\
 R~72 & 0.62 & 0.72 & 209$\pm$ 15 & 210 $\pm$ 20 & 141 $\pm 15$ \\
 R~85 & 0.48 & 0.54 & 103$\pm$ 10 &  90 $\pm$ 10 &  66 $\pm 10$ \\
 R~89 & 1.20 & 1.31 & 265$\pm$  8 & 425 $\pm$ 15 & 260 $\pm 10$ \\
 R~92 & 0.65 & 0.74 & 195$\pm$ 10 & 210 $\pm$  5 & 147 $\pm  6$\\
 R~93 & 1.33 & 1.58 &  60$\pm$ 10 & 547 $\pm$ 17 & 303 $\pm 17$\\
 R~94 & 1.35 & 1.69 & 174$\pm$  8 & 535 $\pm$ 14 & 313 $\pm 15$\\
 R~95 & 0.83 & 0.93 & 317$\pm$ 15 & 350 $\pm$ 10 & 184 $\pm 7$ \\
 R~96 & 1.21 & 1.33 & 320$\pm$ 12 & 469 $\pm$ 18 & 251 $\pm 11$\\
% & &  & & &\\
%\multispan{6}{IC~2391 \hfill}\\
% & &  & & &\\
% VXR~16A & ---  &  0.93 & 302 $\pm 12$ & 325 $\pm 13$ & 184 $\pm 10$ \\
% VXR~18A & ---  &  1.52 &  55 $\pm 15$ & 470 $\pm 30$ & 289 $\pm 13$ \\
% VXR~62A & 0.85 &  0.98 & 250 $\pm 15$ & 330 $\pm 20$ & 190 $\pm 14$ \\
% VXR~67A & ---  &  1.02 & 255 $\pm 16$ & 319 $\pm 30$ & 210 $\pm 15$ \\
% VXR~69A & ---  &  0.89 & 251 $\pm 17$ & 290 $\pm 12$ & 177 $\pm 10$ \\
% VXR~72  & 0.72 &  0.83 & 225 $\pm 13$ & 247 $\pm  7$ & 165 $\pm  7$ \\
% VXR~76A & 1.04 &  1.23 & 204 $\pm 11$ & 382 $\pm 26$ & 241 $\pm 13$ \\
 & &  & & &\\ \hline\hline
\end{tabular}
\end{table*}
\section{Observational data}
The equivalent widths of the K~{\sc i}~7699~\AA~line were
measured in a sample of IC~2602 stars
selected from R97 and R01. 
Ca~{\sc i}~6718~\AA~line strengths were also measured for the sample
stars. Most of the sample stars
had been observed using CASPEC at the ESO 3.6~m telescope; high resolution
spectra for a few of them were acquired using
the Echelle spectrograph on the CTIO 4~m telescope. We refer to R97, R01,
and Stauffer et al. (\cite{sta97})
for details on the observations; we briefly recall here
that the resolving powers ranged between R$\sim 20,000$ and R$\sim 41,000$,
due to the use of different instruments and/or slit widths. 
S/N ratios
in the K~{\sc i}~line spectral region are in the range
between 60 and 100, although a few spectra were characterized by lower
S/N ratios, which did not allow us to measure their potassium features.
The present sample includes
all the stars with \bmo~$\leq 1.4$ for which we were able to get a reliable
measurement of the EW of the K~{\sc i} feature; 
EWs were measured,
as usual, by direct integration below the continuum levels.
These in turn were estimated by polynomial fitting of line free regions.
The sample stars are listed in Table~1.
Star names come from Randich et al. (\cite{R95}) 
(with exception
of W79 --from Whiteoak \cite{whi61});
in Cols.~2 and 3 we list \bmo~and \vmo~colors which were
retrieved from Prosser et al. (\cite{pros96}); as in R01 reddening
values E(\bmv)$=$E(\vmi)$=0.04$ were assumed.
The measured EWs of the Li~{\sc i} (retrieved from R97 and R01), 
K~{\sc i}, and Ca~{\sc i}~lines,
together with $1\sigma$ errors, are listed in Cols.~4--6.
\section{Results}
\subsection{The dispersion in the EW vs. color diagrams}
In Fig.~1 we plot the
equivalent width of the potassium line vs.
\bmo~(left-hand panel) and \vmo~(right-hand panel) colors for IC~2602 
members included in the present sample; 
Figure~2 and Fig.~3 are the same as Fig.~1, but the EWs of the Li~{\sc i}~
and Ca~{\sc i}~6718~\AA~lines are shown, respectively.
It is worthwhile recalling that calcium is not an alkaline element;
however, since the Ca~{\sc i}~6718~\AA~is not thought to be affected by
activity, it is used here for comparison purposes.
\begin{table*}
\caption{The scatter in the EW vs. color diagrams: reduced $\chi^2$ values
and probabilities that the dispersion is real (numbers within parenthesis)
are listed.
``S" means a probability larger than 99.9 \%, while ``NS" indicates
a probability below 67 \%.
``Degree" indicates the degree of the polynomial fitting of the EW vs. color
relationship (see text for details). ``All" refers to the whole color range,
while intervals A, B, and C, refer to \bmo$< 0.75$, 0.75$\leq$\bmo$\leq 1$, and
\bmo$>1$, respectively.}
\small
\begin{tabular}{|c|cccc|cccc|c|}
 \multicolumn{10}{c}{} \\ \hline
% & & & & & & & & &     \\ \hline
 color/degree  & \multicolumn{4}{|c|}{Li~{\sc i}~(6708~\AA)} &\multicolumn{4}{c|}{K~{\sc i}~(7699~\AA)}& Ca~{\sc i}~(6718~\AA)\\
      & All & A & B  & C & All & A & B & C & All \\
 & & & & & & & & &     \\ \hline
\bmv/1 & --- & --- & --- & --- & 12.6 (NS) & 11.3 (NS) & 19.6 (98~\%) & 2.4 (NS) & 1.5 (NS)\\
\bmv/2 & 36.1 (91~\%) & 6.9 (NS) & 66 (S) & 42 (S) & 12.2 (NS) & 10.7 (NS) & 17.4 (96~\%) & 2.9 (NS) & 1.6 (NS) \\
\bmv/3 & 37.2 (94~\%) & 6.8 (NS) & 66 (S) & 41 (S) & 13.1 (NS) & 12.1 (72~\%) & 17.5 (96~\%) & 1.8 (NS) & 1.3 (NS)\\
\bmv/4 & 36.2 (95~\%) & 6.3 (NS) & 60 (S) & 41 (S) & --- & --- & --- & --- &
--- \\
 & & & & & & & & &     \\ \hline
\vmi/1 & --- & --- & --- & --- & 6.7 (NS) & 4.7 (NS) & 4.9 (NS) & 4.1 (NS) & 2.5 (NS)\\
\vmi/2 & 27.8 (NS) & 6.6 (NS) & 42 (S) & 43 (S) & 3.1 (NS) & 1.4 (NS) & 1.3 (NS) & 1.6 (NS) & 2.4 (NS) \\
\vmi/3 & 28.9 (73~\%) & 6.5 (NS) & 41 (S) & 44 (S) & 3.2 (NS) & 1.8 (NS) & 1.3 (NS) & 1.6 (NS) & 2.4 (NS) \\
\vmi/4 & 28.9 (77~\%) & 7.9 (NS) & 50 (S) & 31 (S) & --- & --- & --- & --- & --- \\
 & & & & & & & & &     \\ \hline\hline
\end{tabular}
\normalsize
\end{table*}
The comparison of the three figures first shows that, whereas the EWs
of the K~{\sc i}
and Ca~{\sc i} lines monotonically increase with color, as well known,
this is not
the case for lithium. The difference in the EW vs. color morphologies
of the three lines is due to the fact that, as mentioned in Sect.~1
(and this is indeed the motivation for the present and similar studies),
lithium is destroyed in low mass stars, with the amount of depletion
increasing with decreasing mass,
while potassium and calcium are not;
therefore, the EWs of the Ca~{\sc i} and K~{\sc i} lines are determined by
effective temperature only, since 
we are considering stars in the same evolutionary
status, and thus with similar gravities and microturbulence values; on
the contrary, the
strength of the Li line is determined by both the effective
temperature (for a fixed abundance the EW would increase with
decreasing temperature) and the abundance, which decreases towards cooler stars.
The EW(Li) vs. color morphology for cluster stars is indeed
a very well known result and a more detailed discussion is not warranted
here. It is instead important to stress again that, 
according to standard models,
at a given color or effective temperature, stars in the same cluster
should have the same Li abundance, and thus Li EW,
and a tight Li vs. color distribution should be observed.
\begin{figure*}
\psfig{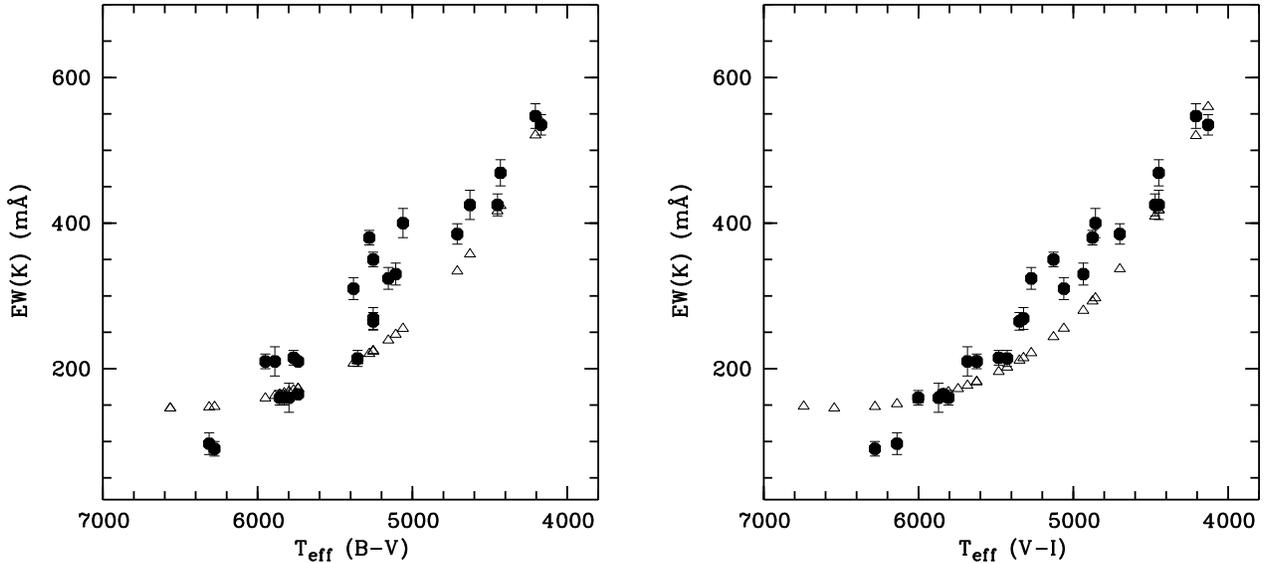}
\vspace{-2cm}
\caption{Potassium EWs vs. \teff~inferred from \bmv~colors
(left-hand panel) and \teff~from \vmi~colors. Filled symbols denote
measured EWs for our sample stars, while open symbols indicate predicted
EWs using the calibration of Tripicchio et al. (\cite{tri99}) (see text).}
\end{figure*}
\begin{figure*}
\psfig{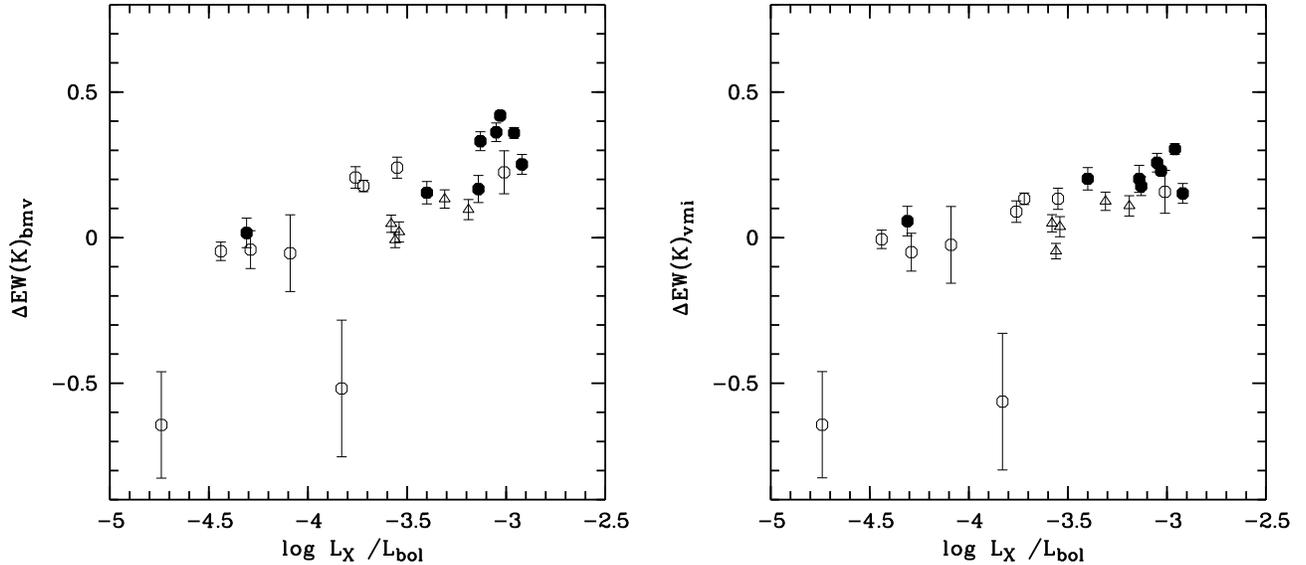}
\vspace{-2cm}
\caption{Differential potassium EWs ($\Delta$EW$=$(EW$_{\rm obs}-
\rm EW_{\rm pred})$/EW$_{\rm obs}$) are plotted as a function of the logarithm
of the ratio of
X-ray over bolometric luminosity, used as an activity
tracer. Left- and right-hand panels 
show differential EWs computed with respect to predicted EWs based on 
\teff(\bmv) and \teff(\vmi), respectively. Open circles denote stars with
\bmo $< 0.75$, filled circles stars with $0.75 \leq$\bmo$\leq 1$, and open
triangles stars with \bmo$>1$. Note that \logra~values are not available
for all our sample stars.}
\end{figure*}
\begin{figure*}
\psfig{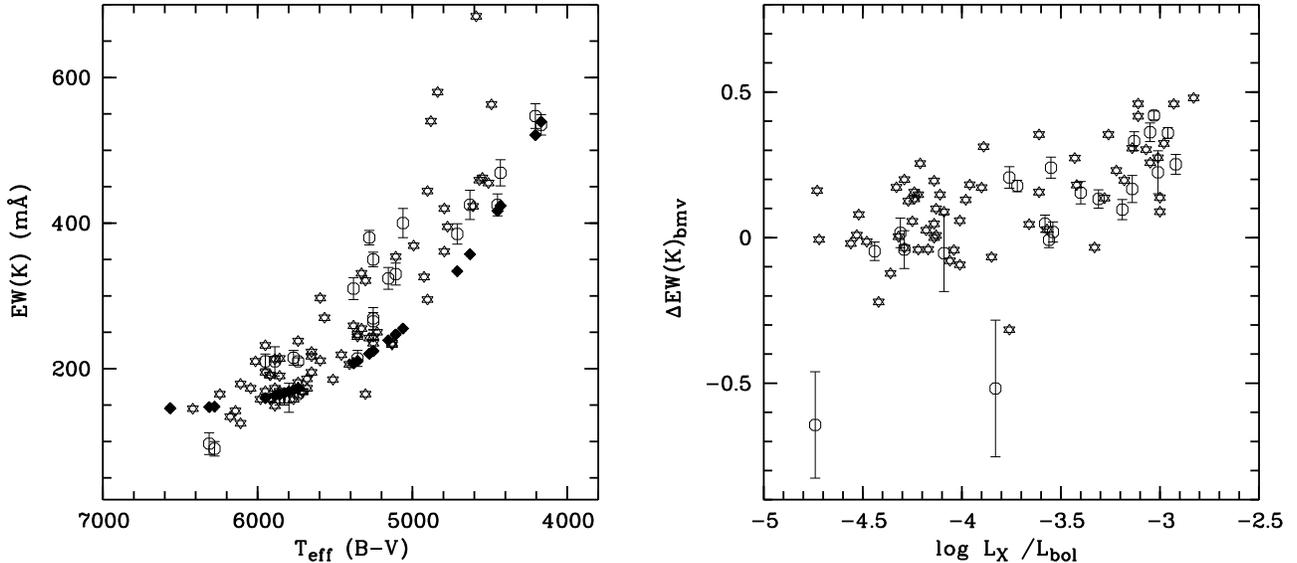}
\vspace{-2cm}
\caption{Left-hand panel: potassium EWs vs. \teff(\bmv). Filled symbols
indicate predicted EWs based on Tripicchio et al. (\cite{tri99}),
while open circles and stars denote the observed EWs for IC~2602
and the Pleiades, respectively. Right-hand panel: \bmv~based differential
potassium EWs as a function of \logra~for our sample stars (open circles)
and the Pleiades (open stars).}
\end{figure*}
Figures~1--3 also 
suggest that: {\it i)} as already pointed out by R97 and R01,
a dispersion in the EW of the Li line is present among cluster stars,
i.e., stars with similar colors have different EWs; {\it ii)}
A dispersion in potassium EWs seems also to be present
in the EW~vs. \bmv~diagram; the dispersion seems
narrower when looking at the EW vs. \vmi~diagram; {\it iii)}
The dispersion is most evident for stars with 0.8 $\la$~\bmo~$\la 1$;
on the contrary, late--K stars (\bmo$\ga 1$) do not show a large
spread in potassium EWs, although they do
show a dispersion in Li (see Fig.~2); {\it iv)}
Finally, no significant (i.e., larger than measurement errors)
dispersion seems to be present in the calcium EW vs. color diagrams.

We carried out a more quantitative analysis
by performing polynomial regressions of the observed EW vs. color
distributions and estimating reduced $\chi^2$ values that allow
assessing on a statistical basis the presence/lack of a star-to-star scatter. 
The reduced $\chi^2$ values 
for the three atoms as a function of the two colors and
for different \bmv~ranges
are listed in Table~2. In the table we also provide the probability
that the observed dispersions are real. The symbol
``S" means a probability larger than 99.9 \% (i.e., a significance level
larger than 3$\sigma$), while ``NS" means a probability
below 67 \% (i.e., below 1$\sigma$ significance).
The regression curves are also shown in the three
figures. Note that the polynomial regressions we have carried
out do not have
any real physical meaning (i.e., we are not trying to model the
EW vs. color patterns); 
our aim here is to infer a ``mean" EW as a function of color and to ascertain
whether the scatter around this mean EW is significant or not.

Table~2 suggests the following points: {\bf 1)} 
The dispersion in lithium
EWs is real in both the EW vs. \bmv~and the EW vs. \vmi~diagrams
for stars in the color ranges B ($0.75 \leq$~\bmo~$\leq 1$) and 
C (\bmo~$>1$). As already known from previous studies,
the dispersion in Li EWs is instead
not significant for earlier-type stars;
{\bf 2)} Our quantitative analysis
supports points {\it ii)} and {\it iii)} above: namely,
the dispersion in the EW(K) vs. \bmv ~diagram
for stars with \bmo~in the range $0.75-1$ is significant, while
the dispersion is not significant
for later and earlier--type stars. The dispersion is also not significant
when considering \vmi~colors.
Note that, as just mentioned in point {\bf 1)},
stars with \bmo $>1$ do show a star-to-star scatter in Li EWs;
{\bf 3)} The EW(Ca) vs. color diagrams are not characterized by a 
significant scatter; we get probabilities larger than 99.9 \%
that the observed dispersion occurs by chance. 
As a final remark, we note that the reduced $\chi^2$ values obtained for the
three lines and, most
important, the probabilities that the observed dispersions are real, 
show a very weak dependence on
the choice of the degree of the fit, i.e., a first order, linear
fit provides similar results than a second order fit (see also Figs.~1--3).
In other words, our results (and in particular the significance of
the dispersion) do not appear to depend on the fitting procedure.
\subsection{Potassium excess vs. activity for the IC~2602 and the Pleiades}
In the previous section we have shown that late--G and early--K IC~2602 stars
are characterized by a dispersion in potassium EWs, at least when considering
the EW(K) vs. \bmv~distribution.
The next issue is whether such a dispersion is caused by activity and,
in particular, whether stars with different activity levels show any
excess/deficit in their potassium EWs.

To address this question we computed for each star a predicted
potassium EW based on the empirical
EW(K) vs. \teff~relationship found by Tripicchio
et al. (\cite{tri99}) for low activity field stars.
More specifically, the predicted
EWs of the potassium line were determined
using equation (1) in Tripicchio
et al. (\cite{tri99}) and the {\it a$_{i}$} coefficients for dwarf stars.
Under the reasonable assumption that our cluster stars have a similar potassium
abundance than field stars, 
differential potassium EWs were then calculated as:\\
$\Delta\rm EW = (\rm EW_{\rm obs.}-\rm EW_{\rm pred.})/\rm EW_{\rm obs.}$,\\
where EW$_{\rm obs.}$ and EW$_{\rm pred.}$ are the observed and predicted
values of the EW, respectively.
Effective temperatures (and thus differential EWs) were derived based
on both the \bmv~and \vmi~colors. \teff(\bmv) were estimated
using a similar \teff~vs. \bmv~calibration than 
Tripicchio et al. (\cite{tri99}),
namely, that of Gray (\cite{gr92}).
Since Gray (\cite{gr92}) does not provide a \teff~vs. \vmi~calibration,
we estimated \teff(\vmi) as follows: we first computed the
difference $\Delta$\teff~between \teff(\bmv) 
from Gray calibration and \teff(\bmv)
from the calibration employed by R01. Then we fitted the relation
between $\Delta$\teff~and \teff(\vmi) from R01 and computed \teff(\vmi)$_
{\rm Gray}=$\teff(\vmi)$_{\rm R01}+\Delta$\teff.
Before presenting and discussing our results, we caution that our
quantitative estimate of differential
potassium EWs or potassium excesses is based on the fit of
Tripicchio et al. (\cite{tri99}) and its accuracy; the comparison
of our measured EWs with predicted EWs from different calibrations
may not necessarily lead to the same quantitative results.

In Fig.~4 we plot the predicted (open triangles) and observed (filled
circles) EWs as a function of \teff(\bmv) (left-hand panel) and
\teff(\vmi) (right-hand panel).
The figure indicates that a large fraction of our sample stars indeed
have larger EWs than predicted; 
stars with \teff~between 5400 and 5000~K (0.75
$\la$~\bmo~$\la 0.95$)
exhibit the largest excesses in the EW vs. \teff(\bmv) diagram; 
however, part of the stars that show an
EW excess in the EW vs. \teff(\bmv) diagram, have an EW more in agreement with
the predicted value when considering the EW vs. \teff(\vmi) diagram.

In Fig.~5 we plot the differential potassium EWs vs. $\log \rm L_{\rm X}/
\rm L_{\rm bol}$,
the ratio of the X-ray over bolometric luminosity.
L$_{\rm X}/\rm L_{\rm bol}$ values were retrieved
from Stauffer et al. (\cite{sta97}) and are used here as activity tracers.
If we exclude the two datapoints with the lowest differential
equivalent widths\footnote{These datapoints correspond to the two 
warmest stars in the sample; they have rather weak potassium EWs and
we cannot exclude that their large deficit in EW is due to measurement
errors.}, a correlation between \logra~values and differential
EWs is evident in both panels. Although a one-to-one relationship
between the two quantities cannot be claimed and for each \logra~value a certain
amount of scatter in $\Delta$EW is present,
stars with larger activity generally
have larger $\Delta$EW values, or larger excesses in EW. We computed 
the one-sided correlation coefficients finding that the correlations 
in both panels are
significant at a confidence level larger than 99.99 (i.e., $> 5\sigma$),
quantitatively
confirming that more active stars tend to have larger \ewk.
We note that whereas a relationship between activity and EW excess
may be present also for our sample stars with \bmo $\geq 1$,	
these stars cover a rather narrow range
of \logra~values and, consequently, a
narrow range of differential EWs; finally, for a given \logra~or activity 
level, several stars have larger differential EWs when considering
the predicted EW based on \teff(\bmv) than those based on \teff(\vmi).

Figures~6a and 6b are similar to the left-hand panels of Figs.~4 and 5, 
but our sample stars are compared to the Pleiades. Potassium 
EWs for this cluster were retrieved from
Soderblom et al. (\cite{sod93a}), while $\log \rm L_{\rm X}/\rm L_{\rm bol}$
values were taken from Stauffer et al. (\cite{sta94}) and Micela
et al. (\cite{mic96}). As for our sample stars, effective temperatures were
inferred using the \teff~vs.~\bmv~ calibration of 
Gray ({\cite{gr92}). The figure clearly shows that the two clusters behave
very similarly; in particular Fig.~6b indicates that, above \logra~$\sim
-4$, stars with similar
activity levels have similar excesses in potassium EWs. Low activity
Pleiades stars instead exhibit a scatter in
differential EWs, 
while all IC~2602 members with \logra~$<-4$ have $\Delta$~EW(K)~$\sim 0$.
The rather small number of low-activity IC~2602 members and 
the lack of available errors for the potassium EWs of the Pleiades, 
do not allow us to ascertain
whether this difference is significant or not. In any case, as for IC~2602,
we find that the correlation between differential EWs and \logra~values
for the Pleiades is significant at $> 99$~\% confidence level.
Note that, whereas a relationship between differential EWs and activity
for the Pleiades was already found by King et al. (\cite{king00}), 
we computed EW excesses with respect to inactive field stars, while
their EW excesses (deficits) referred to a mean trend in the EW vs.
\teff~diagrams.

\section{Discussion}
What do the results presented in the previous section allow us to conclude?
First, the findings of King et al. (\cite{king00})
and of similar papers for the Pleiades seem to hold also for the younger
IC~2602: a dispersion in potassium EWs is detected. In addition,
active stars in IC~2602 and the Pleiades show potassium EWs in excess of
those of inactive field stars with similar colors; 
a statistically significant correlation between EW excess
and activity is found for both clusters.
These results support the hypothesis that the appearance
of the alkali EW vs. color distributions
in young clusters is affected by activity which
alters the formation of the resonance lines.
Activity seems to affect stars at all colors, although different dispersions
in the activity levels reflect into different spreads in the potassium EW
for stars in different color ranges.

Second, no significant dispersion in the EW(Ca) vs. color
diagrams is found:
so far no study has tried to model how activity and the
presence of surface inhomogeneities would affect the
observed  Ca~{\sc i} vs. color diagrams.
Activity is not thought to affect the formation of the Ca~{\sc i}
line itself; however, large and cool spots, besides affecting the
observed EWs of the alkali atoms, are also predicted to alter stellar colors 
(Stuik et al. \cite{stu97}, Barrado y Navascu\'es et al. \cite{bar01}).
The Ca EW vs. color distributions
can thus be used to put some constraints on the characteristics of the spots.
If \bmv~colors of our cluster stars were
significantly altered by the presence of surface 
inhomogeneities, a dispersion should also be observed in the EW(Ca) vs. color
diagrams, since stars with the same intrinsic color, 
but different activity levels
would have the same Ca EW, but different observed colors. Viceversa,
at a given observed color one would find stars with different intrinsic colors
and thus Ca EWs. 
More specifically, the absence of a detectable scatter in the EW of the Ca line,
suggests that the characteristics of surface inhomogeneities in
our sample stars should be such that stellar colors are not greatly 
altered. Given the EW(Ca) vs. \bmv~relationship for our sample stars,
$\Delta$\bmv~larger than 0.1~dex are required to have $\Delta$EW(Ca) $> 2\sigma$
(EW(Ca)) and hence a detectable spread in Ca EWs. 

Figures 1b) and 3a) of Barrado y Navascu\'es et al. (\cite{bar01}) 
show the predicted
variations in \bmv~colors and potassium equivalent widths vs. 
spot coverage for different spot 
temperatures. 
Fig.~1b indicates that in order to have variations in \bmv~colors
$\Delta$\bmv~values larger than $\sim 0.1$~dex,
filling factors larger than 40 \% (or more, depending on
the $\Delta$~T between the spot and the quiet photosphere) are needed. 
On the other hand, differential EWs of our sample stars 
can be as large as $\sim 0.4$~dex (see Fig.~5):
the comparison with Fig.~3a of Barrado
y Navascu\'es et al. (\cite{bar01}) shows that, assuming spot coverages below
$\sim 40$ \%, these values can be obtained for  $\Delta$T between the
spot and the quiet photosphere of the order of $\sim 1000$~K.
In other words, the whole range of
differential potassium EWs measured among IC~2602 stars is consistent
with the predictions of Barrado y Navascu\'es et al. (\cite{bar01}),
provided that our sample stars are covered by cool enough spots.

Third, we find that the EW(K) vs. \vmi~diagram is not characterized by
a significant scatter (neither for stars in different color ranges, nor when
considering the whole range); in addition, differential EWs from a
\vmi~based analysis are somewhat
smaller than those from a \bmv~based analysis. This means that most
of our sample stars appear redder/cooler when considering \vmi~colors.
We do not have a definitive explanation for this finding, but can
attempt two different hypothesis: namely,
either \bmv~colors are not correct and thus they are not good temperature
indicators and \vmi~colors should always be used
(but in this case it would be hard to explain the rather tight EW(Ca) vs.
\bmv~relationship);
or \vmi~colors are somewhat more affected by spots than \bmv~colors; if
the increase in EW is accompanied by
a shift in color, an active star with a given color and EW
may move, in the EW(K) vs. color plane, close to a cooler less active star 
with an intrinsically redder color and larger EW. Although Barrado y
Navascu\'es et al. (\cite{bar01}) and similar studies have not modeled
the effect of spots on \vmi~colors it is reasonable to think that cool
enough spots would affect more \vmi~ than \bmv~colors.
\section{Conclusion: is the spread in lithium due to activity ?}
We have measured the EW of the K~{\sc i}~line in a sample of late-type
members of IC~2602.
Our study confirms the results of previous papers based on the Pleiades
cluster, but, at the same
time, adds new pieces of information into the issue of the scatter
in Li/K observed among late-type stars in young clusters.
The results and discussion presented in the previous 
sections support the idea that the potassium EW~vs. color diagrams
of IC~2602 stars are affected by activity. Based on a statistical
analysis, a star-to-star scatter in potassium
EWs is detected in the EW vs. \bmv~diagram of late--G/early--K--type
stars. In addition, the most active IC~2602
(and Pleiades) stars show
EW excesses with respect to inactive field stars; differential EWs as large
as 0.4~dex are measured, with more
active stars generally having larger EW excesses. 
More specifically, a significant correlation between potassium excess
and \logra~ratios is found for both IC~2602 and the Pleiades.
The dispersion in the EW(K) vs.~\vmi~diagram is
not significant and we suggest that this may be due to the different
effect that cool spots have on \bmv~and \vmi~colors.
We also find that stars later than \bmo~$\sim 1$ do not show a significant
dispersion in EW(K), although
they do show EW(K) excesses. We believe that the reason for this
is the narrow range in activity levels covered by late--type cluster
members. 

The question remains whether 
the observed scatter in Li EWs can then be {\it fully} explained
as due to activity. To answer this question 
the same type of analysis that we presented for potassium should in
principle be carried out also for lithium; namely,
the Li EWs of cluster stars should be compared with those of a sample
of inactive field stars in order to look for EW(Li) excesses. 
However, due to the age--magnetic activity inverse
relationship, inactive field stars are most likely old and 
have hence undergone a significant amount of Li depletion; their intrinsic
Li abundance
is presumably lower than that of IC~2602 members and the comparison
would be meaningless.
The comparison of measured EWs in IC~2602 with
model predictions for the age of the cluster would not work either:
standard models depend on the adopted assumptions
on e.g. convective treatment and atmospheric opacities, and different
groups make different quantitative predictions on Li depletion as a function
of mass; therefore the comparison,
and in particular any Li excess/deficit that we would find, would
depend on the particular choice of the model. 

The most secure conclusion that can be drawn from the present study
is that the scatter in lithium
among stars later than \bmo~$\sim 1$ cannot be explained by activity only.
The detection of EW(K) excesses among late--K cluster members
suggests that the formation of the K~{\sc i} feature in these stars
is affected by activity; however, as we have discussed, these stars
(at least those included in the present sample) are characterized by similar
activity levels and thus show little scatter in their potassium EWs.
The same should hold for the Li line, in contradiction with the
large observed scatter in Li among our sample stars with \bmo~$> 1$.
In other words, although the Li EWs of the coolest stars in our
sample are most likely affected by the presence of surface inhomogeneities
(since potassium EWs are affected), the lack of scatter in potassium EWs among
these stars
suggests that the ultimate reason for their dispersion in Li abundances 
is a different amount of Li destruction.

As to earlier-type stars, at this stage we cannot answer the question
whether their scatter in Li EWs is due to a scatter in abundances,
and hence different amounts of
Li depletion, or is instead {\it completely} due to activity.
Certainly, it is at least {\it in part} caused by activity. If we focus
on a narrow color range (e.g. $0.79 \la$~\bmo~$\la 0.84$), stars with
larger potassium EW excesses tend to have larger Li EWs; since within 0.05
dex difference in \bmv~(or $\sim 100$~K interval in \teff) a large
difference in Li abundance is not expected, this provides an hint that
the difference in Li EWs is really due to a difference in activity
only. Unfortunately, there are only five stars in this color range, which
does not allow us to regard this conclusion as definitive.

Most obviously, further investigations on this topic should be carried out,
both on theoretical and observational grounds.
The potassium line should be measured in additional clusters. 
Additional photometry, and in particular photometric
monitoring and/or, as mentioned by Jeffries (\cite{jeff99}), Doppler imaging of 
cluster stars, which would allow constraining the characteristics of
spots and plages and their timescales, should also be performed.
If possible, simultaneous monitoring
of the alkali lines EWs (including also the Na~{\sc i} 5896~\AA~feature)
should be obtained.
At the same time, additional modeling should
be carried out. 
\begin{acknowledgements}
I thank an anonymous referee for her/his very useful comments
and suggestions. I am grateful to John Stauffer for making available to
me CTIO spectra of three stars in the sample.
\end{acknowledgements}
{}
\end{document}